\documentclass[seceq]{ptptex}

\title{
Hilltop Supernatural Inflation%
}

\author{
Chia-Min \textsc{Lin}%
}

\inst{
Department of Physics, National Tsing Hua University, Hsinchu, Taiwan 300
}

\abst{
In this talk, I will explain how to reduce the spectral index to be $n_s=0.96$ for supernatural inflation. I will also show the constraint to the reheating temperature from Big Bang Nucleosynthesis of both thermal and non-thermal gravitino production.
}

\usepackage{graphicx,color}
\usepackage{subfigure}
\begin{document}

\maketitle

\section{Introduction}
Inflation (for the general review, \cite{textbook}) solves many fine tuning problems of conventional hot big bang model and provides the seeds of structure formation.
The spectrum of the primordial curvature perturbation is given by
\begin{equation}
P_R=\frac{1}{12\pi^2M_P^6}\frac{V^3}{V'^2} \;.
\label{spectrum}
\end{equation}
The number of e-folds is given by
\begin{equation}
N=M^{-2}_P\int^{\phi(N)}_{\phi_{end}}\frac{V}{V'}d\phi.
\label{efolds}
\end{equation}
The spectrum corresponds to the CMB scale which we can observe left horizon at $N \sim 60$ and the spectrum is normalized to be $P_R^{1/2} = 5 \times 10^{-5}$ by the Cosmic Microwave Background (CMB) temperature fluctuations. The latest WMAP result \cite{Komatsu:2010fb} suggests a red spectrum with the spectral index $n_s \sim 0.96$. Future experiments like PLANCK satellite may further confirm this result.

This talk is organized as follows. In section~\ref{1}, I introduce the idea of hybrid inflation. In section~\ref{2}, a particular realization of hybrid inflation based on supersymmetry called supernatural inflation is introduced. I also explain the current problem of the model compared with experiments. In section~\ref{3}, I explain the idea of hilltop supernatural inflation and how to realize it. In section~\ref{4}, I will show the constraint to the reheating temperature as a function of gravitino mass. Section~\ref{5} is my conclusion.

\section{Hybrid Inflation}
\label{1}
The potential of the inflaton field for chaotic inflation is
\begin{equation}
V(\phi)=\frac{1}{2}m^2 \phi^2.
\end{equation}
It is widely known that for this model to work, we need a field value $\phi > M_P$ at $N=60$. However, we can reduce the field value to be $\phi < M_P$ by adding a constant term $V_0$ to the potential. The potential becomes
\begin{equation}
V(\phi)=V_0+\frac{1}{2}m^2 \phi^2.
\end{equation}
For this model, naively inflation can never end because of the constant term. However, we can actually release $V_0$ after inflation by a so called "waterfall field". This kind of set up is called hybrid inflation.
\section{Supernatural Inflation}
\label{2}
The idea of supernatural inflation \cite{Randall:1995dj} is that we build a hybrid inflation by using parameters coming from supersymmetry (SUSY) breaking sector. In particular, we consider gravity mediated SUSY breaking. In this case, the SUSY breaking scale is $M_S \sim 10^{11}$ GeV and the soft mass is $m \sim O(1)$ TeV. The scalar potential of the inflaton field $\phi$ during inflation for supernatural inflation is given by
\begin{equation}
V(\phi)=M_S^4+\frac{1}{2}m^2 \phi^2.
\end{equation}
This model can work and satisfy the CMB normalization (that is, $P_R^{1/2}=5 \times 10^{-5}$) naturally "with no (very) small parameters"\footnote{This is in the title of \cite{Randall:1995dj}}.

However, because the potential is concave upward, the predicted spectral index is $n_s>1$ which is not favored by recent WMAP result.
\section{Hilltop Supernatural Inflation}
\label{3}
\begin{figure}[htbp]
 \begin{center}
 \includegraphics[width=0.45\textwidth]{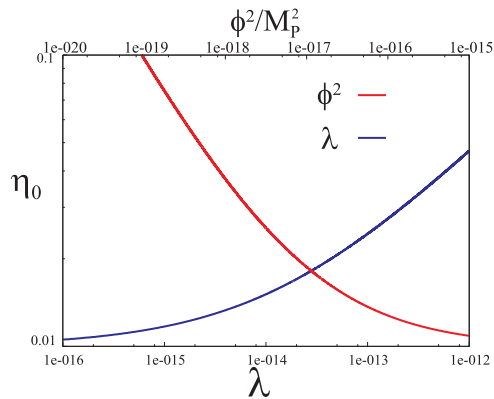}
 \caption{$\phi^2/M_P^2$ and $\lambda$ as a function of $\eta_0$ at $N=60$.}
 \label{result}
 \end{center}
 \end{figure}

The idea of using hilltop inflation \cite{Boubekeur:2005zm, Kohri:2007gq} to reduce the spectral index of supernatural inflation is to add a negative quartic term which naturally exists if we consider a non-renormalizable superpotential:
\begin{equation}
W=\lambda_4 \frac{\phi^4}{M_P}
\label{eq1}
\end{equation}
where $\lambda_4 \sim O(1)$. Therefore the potential we consider is of the following form \cite{Lin:2009yt}
\begin{eqnarray}
V(\phi)&=&V_0+\frac{1}{2}m^2\phi^2-\frac{\lambda_4 A \phi^4}{4M_P}\\
       &\equiv& V_0 \left(1+\frac{1}{2}\eta_0\frac{\phi^2}{M_P^2}\right)-\lambda \phi^4
\label{potential}
\end{eqnarray}
with
\begin{equation}
\eta_0 \equiv \frac{m^2 M_P^2}{V_0}  \;\;\; \mbox{and} \;\;\; \lambda \equiv \frac{\lambda_4 A}{4 M_P}
\end{equation}
where $A \sim O(1)$ TeV. Therefore $\lambda \sim O(10^{-15})$. We choose $V_0=M_I^4$ where $M_I \sim 10^{11}\mbox{ GeV} \sim 10^{-7}M_P$ is the intermediate scale and $\eta_0 \sim O(10^{-2})$. By using Eq.~(\ref{efolds}) and (\ref{spectrum}), we can obtain

\begin{eqnarray}
\left(\frac{\phi}{M_P}\right)^2&=&\left(\frac{V_0}{M_P^4}\right)
\frac{\eta_0 e^{2N\eta_0}}{\eta_0 x+4 \lambda (e^{2N\eta_0}-1)}\\
x &
\equiv & \left(\frac{V_0}{M_P^4}\right) \left(\frac{M_P}{\phi_{end}}\right)^2,
\end{eqnarray}
and
\begin{eqnarray}
P_R&=&\frac{1}{12\pi^2}e^{-2N\eta_0}\frac{[4\lambda(e^{2N\eta_0}-1)+\eta_0 x]^3}{\eta_0^3(\eta_0 x-4\lambda)^2}\\
n_s&=&1+2\eta_0 \left[1-\frac{12\lambda e^{2N\eta_0}}{\eta_0 x+4\lambda(e^{2N\eta_0}-1)}\right].
\end{eqnarray}
After imposing the CMB normalization ($P^{1/2}_R=5 \times 10^{-5}$) and requiring $n_s=0.96$. We can solve $\phi$ and $\lambda$ as a function of $\eta_0$. The result is shown in Fig.~\ref{result}.
As we can see from the plot, we can successfully reduce the spectral index to $n_s=0.96$ without tuning the parameters. The reason I show the field value of $\phi$ is that as long as $\phi$ is small enough, we can safely ignore the contribution from the non-renormalizable F-term coming from Eq.~(\ref{eq1}). For more detail, please see \cite{Lin:2009yt}.

\section{Gravitino Problem}
\label{4}
\begin{figure}[htbp]
\begin{center}
\includegraphics[width=0.5\textwidth]{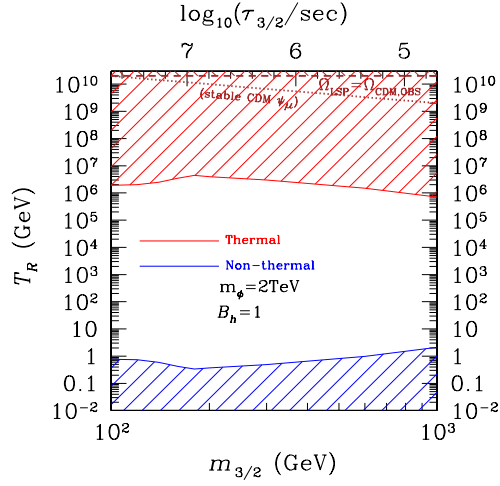}
\caption{Allowed  region in parameter space of $T_R$ versus $m_{3/2}$ with $m_{\phi}=2\mbox{ TeV}$. Note that the constraint can be much
milder only at around $m_{\phi} \sim 2 m_{3/2}$ because of the
suppression of the mode decaying into two gravitinos.}
\label{fig1}
\end{center}
\end{figure}

\begin{figure}[htbp]
\begin{center}
    \includegraphics[width=0.5\textwidth]{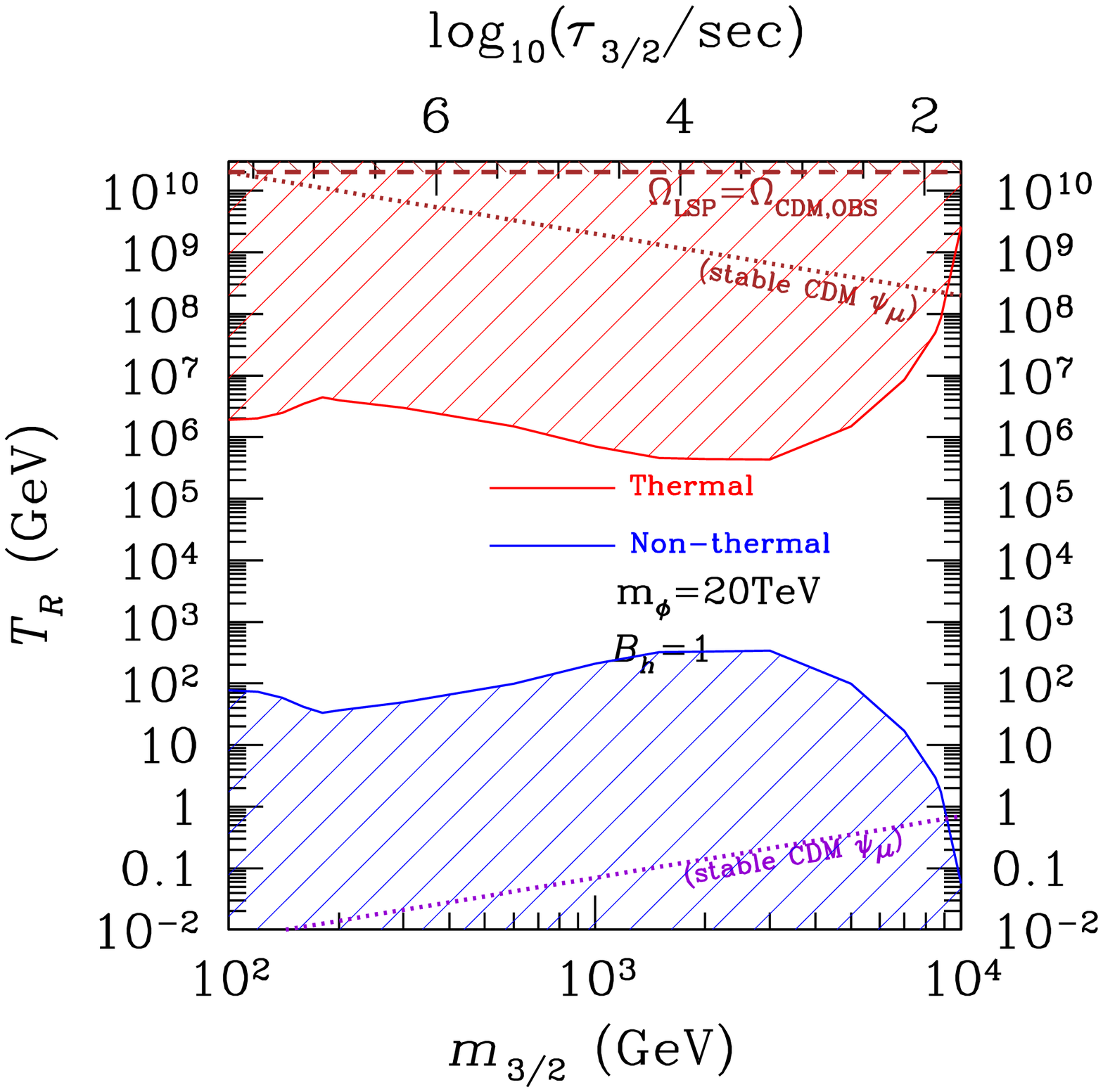}
\caption{$T_R$ versus $m_{3/2}$ with $m_{\phi}=20\mbox{ TeV}$}
\label{fig2}
\end{center}
\end{figure}

\begin{figure}[htbp]
\begin{center}
\includegraphics[width=0.5\textwidth]{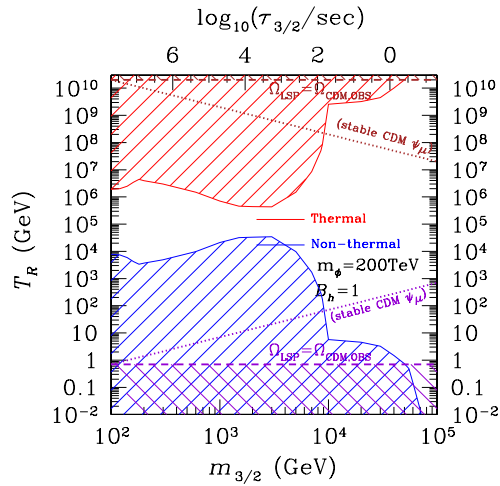}
\caption{$T_R$ versus $m_{3/2}$ with $m_{\phi}=200\mbox{ TeV}$}
\label{fig3}
\end{center}
\end{figure}

One problem of building an inflation model based on SUSY is that it is possible to suffer the so called gravitino problem. Gravitino can be produced either thermally or non-thermally. Thermal production of gravitino would impose an upper bound to the reheating temperature \cite{Kawasaki:1994af}. Non-thermal production of gravitino would impose a lower bound \cite{Kawasaki:2006gs}.
We consider both constraints in \cite{Kohri:2010sj}. The results is the constraint to the reheating temperature as a function of gravitino mass and is shown in Figs.~\ref{fig1},\ref{fig2},\ref{fig3}.

\section{Conclusion}
\label{5}
In this talk I have already shown that it is possible to reduce the spectral index to the value suggested by current observation. We assume SUSY breaking is gravity mediated. The cosmological consequences of gauge mediation has been considered in another work \cite{Lin:2009ux}.

\section*{Acknowledgements}
This is a proceeding for the YKIS2010 conference held in YITP at Kyoto University.
I would like to thank the organizers of the conference for supporting me and gave me the opportunity to present my work.

%

\end{document}